\documentclass[preprint2]{aastex6}

\usepackage{epstopdf}
\usepackage{amsmath}
\begin{document}

\title{Evidence for a CO desorption front in the outer AS 209 disk}
\shorttitle{CO desorption in the AS 209 disk}

\author{Jane Huang, Karin I. \"Oberg, and Sean M. Andrews}

\affil{Harvard-Smithsonian Center for Astrophysics, 60 Garden St., Cambridge, MA 02138, USA; \href{mailto:jane.huang@cfa.harvard.edu}{jane.huang@cfa.harvard.edu}}

\begin{abstract}
Millimeter observations of CO isotopologues are often used to make inferences about protoplanetary disk gas density and temperature structures. The accuracy of these estimates depends on our understanding of CO freezeout and desorption from dust grains. Most models of these processes indicate that CO column density decreases monotonically with distance from the central star due to a decrease in gas density and freezeout beyond the CO snowline. We present ALMA Cycle 2 observations of $^{12}$CO, $^{13}$CO, and C$^{18}$O $J=2-1$ emission that instead suggest CO enhancement in the outer disk of T Tauri star AS 209. Most notably, the C$^{18}$O emission consists of a central peak and a ring at a radius of $\sim1''$ (120 AU), well outside the expected CO snowline. We propose that the ring arises from the onset of CO desorption near the edge of the millimeter dust disk. CO desorption exterior to a CO snowline may occur via non-thermal processes involving cosmic rays or high-energy photons, or via a radial thermal inversion arising from dust migration.     
\end{abstract}


\section{Introduction}
Characterizing protoplanetary disk structures is essential for understanding planet formation. Disk properties are often inferred by modeling millimeter observations of CO isotopologues (e.g.  \citealt{2003AA...399..773D,2013ApJ...774...16R, 2014ApJ...788...59W}). The vertical structure of a disk is typically described in terms of a surface layer in which strong UV radiation photodissociates molecules, an intermediate warm molecular layer, and a cold midplane \citep{2002AA...386..622A}. The relationship between CO abundances and overall gas density structure is tied to the disk temperature profile. Near the star, CO is in the gas phase, but in the cold outer disk midplane, CO is expected to be frozen out onto dust grains. The boundary separating these two regions is the CO snowline, thought to influence the formation locations and compositions of planets and planetesimals  (e.g., \citealt{2011ApJ...743L..16O, 2013Sci...341..630Q,2014ApJ...793....9A}).  While the fractional abundance of gas-phase CO is generally assumed to be constant in the warm molecular layer, depletion due to photodissociation and freezeout in other layers must be accounted for in order to avoid underestimating the disk mass or surface density \citep{2011ApJ...740...84Q}. Because $^{12}$CO is optically thick in disks, $^{13}$CO and C$^{18}$O observations have been used to derive disk masses, but these may be substantially underestimated if selective photodissociation is neglected \citep{2014AA...572A..96M}. Comparisons of gas mass estimates  from HD and C$^{18}$O observations for the TW Hya disk also indicate that the CO:H$_2$ ratio in the warm molecular layer may not follow typical ISM abundances \citep{2013Natur.493..644B, 2013ApJ...776L..38F}. 

Desorption in the outer disk midplane, preventing the complete freezeout of CO exterior to the snowline, may be an additional complicating factor in characterizing disk structure with CO observations. Two mechanisms may return CO to the gas phase in the outer disk. First, non-thermal desorption may occur when disk densities decrease enough for cosmic rays, X-rays, and UV radiation to penetrate to the midplane (e.g. \citealt{2007ApJ...660..441W, 2010ApJ...722.1607W}). Second, the inward drift of large grains or external photoevaporation may invert the radial temperature profile, leading to thermal CO desorption \citep{2016ApJ...816L..21C, 2016MNRAS.457.3593F}. {\"O}berg et al.  \citeyearpar{2015ApJ...810..112O} presented indirect evidence for CO desorption in the outer disk of T Tauri star IM Lup with a detection of double DCO$^+$ rings, attributing the outer ring to increased production of DCO$^+$  due to UV photodesorption of CO ice. 

In this letter, we present evidence for a CO desorption front in the outer disk of AS 209, based primarily on the emission pattern of C$^{18}$O. In sections 2 and 3, we describe ALMA observations of $^{12}$CO, $^{13}$CO, and C$^{18}$O $J=2-1$ transitions in the disk around AS 209, a 1.6 Myr T Tauri star with a mass of 0.9 $M_\odot$ \citep{2009ApJ...700.1502A}. The system is thought to be part of the Ophiuchus star-forming region about 120 pc away \citep{2008ApJ...675L..29L}. In section 4, we simulate observations with a toy model to demonstrate that a CO abundance enhancement in the outer disk is consistent with the different isotopologue emission morphologies. Section 5 discusses approaches for further characterizing CO desorption in disks and implications for disk structure inference. 

\section{Observations and Data Reduction}
AS 209 (J2000.0 R.A. $16^{\textup{h}}49^{\textup{m}}15^{\textup{s}}.29$, decl. $-14\degr 22'08\farcs 6$)  was observed with the Atacama Large Millimeter/Submillimeter Array on 2014 July 2 (project code ADS/JAO.ALMA\#2013.1.00226), with 21 minutes on source. The configuration consisted of thirty-four 12 m antennae, with baselines between 20 and 650 m. Observations were set up with thirteen Band 6 spectral windows (SPWs). Twelve (including those with CO isotopologue lines) had spectral resolutions of 61 kHz and bandwidths of 59 MHz, while the thirteenth had a resolution of 122 kHz and bandwidth of 469 MHz. An earlier reduction of the 1.4 mm continuum and $^{13}$CO data was published in \citet*{2015ApJ...809L..26H} as part of an N$_2$D$^+$ analysis.  

ALMA/NAASC staff provided a calibration script using the quasar J1733-1304 for bandpass and phase calibration and Titan for flux calibration. It was modified to scale visibility weights properly\footnote{See \newline\url{https://casaguides.nrao.edu/index.php/DataWeightsAndCombination}} and executed in CASA 4.4.0 to calibrate visibilities. The CO isotopologue SPWs were phase self-calibrated with solutions obtained from averaging six SPWs free of strong line emission. After subtracting the continuum in the uv-plane, each line was imaged and CLEANed with Keplerian rotation masks. The 1.4 mm dust continuum noise level is $\sigma =0.27$ mJy beam$^{-1}$, and the ALMA systematic flux uncertainty is $\sim$10$\%$. The continuum flux density, obtained by integrating interior to the 3$\sigma$ contour, is 252 $\pm$ 25 mJy, which is consistent with Submillimeter Array observations of AS 209 presented in  \citet{2011ApJ...734...98O}. Channel rms for 0.1 km s$^{-1}$ bins and integrated flux values are listed in Table \ref{Table1}. 

\begin{deluxetable*}{cccccc}
\tablecaption{Summary of Line Observations\label{Table1}}
\tablehead{
\colhead{Transition} & \colhead{Rest Frequency} &\colhead{ E$_u$} &\colhead{ Beam (P. A.)} &\colhead{Channel rms}&\colhead{Integrated Flux}\\
\colhead{}& \colhead{(GHz)} & \colhead{(K)} &\colhead{}&\colhead{(mJy beam$^{-1}$)}& \colhead{(Jy km s$^{-1}$) }
}

\startdata
$ ^{12}$CO $J = 2-1$ & 230.53800&16.6 &$0\farcs 62\times0\farcs 57$ $(-78\fdg 48)$ &10&7.5$\pm$0.8 \\
$^{13}$CO $J = 2-1$ & 220.39868 & 15.9 &$0\farcs 63\times 0\farcs 58$ $(74\fdg 31)$ &15&2.1$\pm$0.2\\
 C$^{18}$O $J = 2-1$ &219.56035 &15.8  &$0\farcs 65\times 0\farcs 59$ $(-72 \fdg 99)$ &8 &0.54$\pm$0.05\\
\enddata

\end{deluxetable*}

\section{Observational Results}
\begin{figure}[htp]
\epsscale{1.3}
\plotone{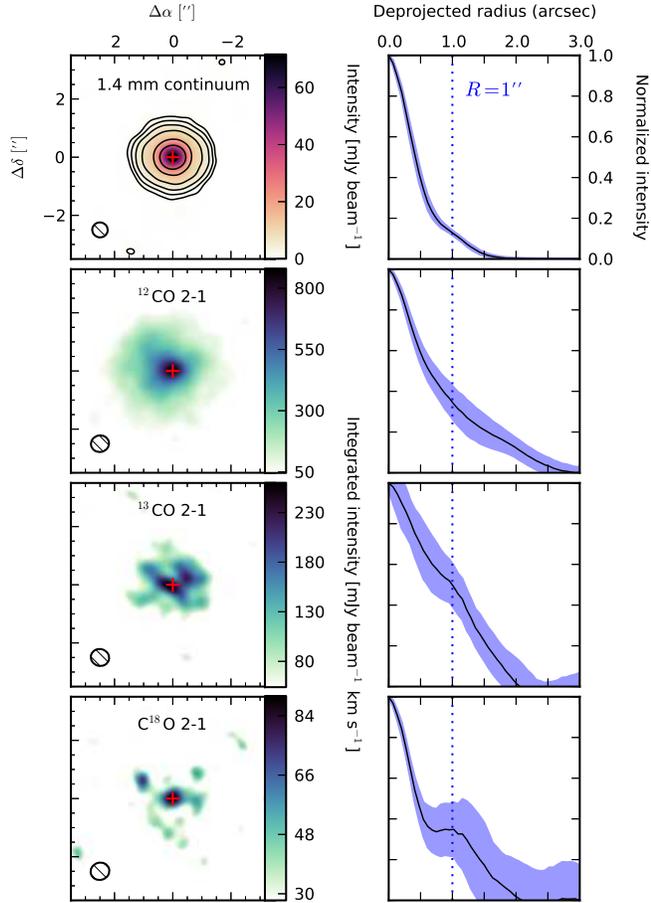}
\caption{AS 209 disk emission maps and radial intensity profiles. \textit{Left column}: Top left panel shows the 1.4 mm continuum intensity, with [4, 8, 16, 32...]$\sigma$ contours, where $\sigma= 0.27$ mJy beam$^{-1}$. The next three panels are $^{12}$CO, $^{13}$CO, and C$^{18}$O $J=2-1$ integrated intensity maps. Line emission color bars start at 2$\sigma$. Synthesized beams are shown in each panel's lower left. Red crosses mark the continuum peak position. \textit{Right column}: Deprojected and normalized continuum, $^{12}$CO, $^{13}$CO and C$^{18}$O $J=2-1$ radial intensity profiles. Purple ribbons show the standard deviation at each radius.}
\label{fig1}
\end{figure}

\begin{figure}[htp]
\epsscale{1.2}
\plotone{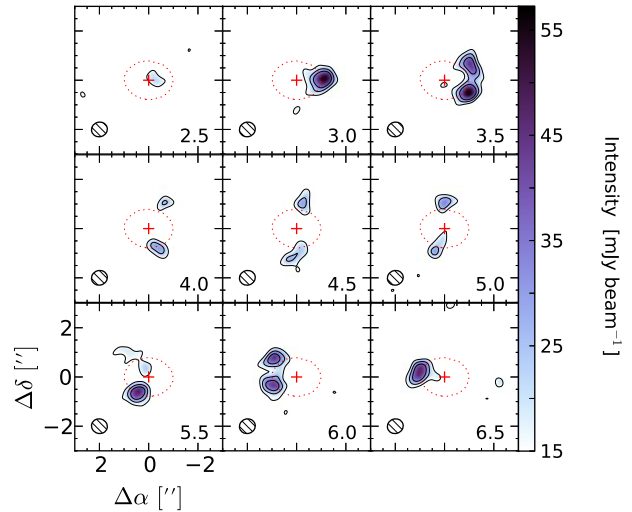}
\caption{Channel maps for the AS 209 disk C$^{18}$O $J=2-1$ transition. Synthesized beams are shown in the lower left and channel velocities  (km s$^{-1}$)  in the lower right  of each panel. Red crosses mark the phase center. The red ellipse traces a $1''$ projected radius, assuming P.A. = 86$\degr$  and inclination = 38$\degr$. Contours are [3, 5, 7]$\sigma$, with $\sigma$ = 5 mJy beam$^{-1}$. 
\label{fig2}}
\end{figure}

Column one of Figure \ref{fig1} shows the 1.4 mm dust continuum intensity and $^{12}$CO, $^{13}$CO, and C$^{18}$O $J=2-1$ integrated intensity maps (summed from -2.0 to 11.0 km s$^{-1}$). Column two of Figure \ref{fig1} shows corresponding deprojected and azimuthally averaged intensity profiles. The position angle and inclination of 86$\degr$ and 38$\degr$, respectively, are adopted from \citealt{2009ApJ...700.1502A}. The continuum intensity is normalized to its peak value, 71.7 mJy beam$^{-1}$ (synthesized beam: $0\farcs 53\times0\farcs 51$ $(-78\fdg 07)$). Line profiles are normalized to the peak integrated intensities of 881, 266, and 91 mJy beam$^{-1}$ km s$^{-1}$  for  $^{12}$CO, $^{13}$CO, and C$^{18}$O, respectively. The rms in the integrated intensity maps are $\sigma = $ 23, 27, and 14 mJy beam$^{-1}$ km s$^{-1}$ for $^{12}$CO, $^{13}$CO, and C$^{18}$O, respectively. 

The continuum and $^{12}$CO emission are centrally peaked and decrease monotonically with radius, although there is significant $^{12}$CO emission outside the continuum detection threshold. The $^{12}$CO emission is weaker on the west side, likely due to cloud contamination \citep{2011ApJ...734...98O}. In contrast to $^{12}$CO, the $^{13}$CO profile bulges outward at $\sim1''$ (120 AU). Like the other isotopologues, the C$^{18}$O emission is centrally peaked, but also has a ring at $\sim1''$ coinciding with the $^{13}$CO ``bulge."  

Figure \ref{fig2} shows C$^{18}$O $J=2-1$ channel maps. A red ellipse traces a $1''$ projected radius. Line wings contribute to the central peak observed in the integrated intensity map, but in channels near the systemic velocity of $\sim 4.5$ km s$^{-1}$, little emission is present within $1''$ of the disk center. Like the C$^{18}$O integrated intensity map, the channel maps demonstrate that the emission traces out a ring at $\sim1''$. 

\section{A toy model for CO desorption in the outer disk}
A CO desorption front, the onset of desorption in the outer disk outside a CO snowline, may produce an emission ring. To explore qualitatively how such a CO distribution can affect the emission profiles of different isotopologues, we simulated observations of a T Tauri disk using a toy model adapted from parametric gas density and temperature models that have been applied to CO emission in many disks (e.g. \citealt{2013ApJ...774...16R,2014ApJ...788...59W,2015ApJ...806..154C}). Of the isotopologues observed in the AS 209 disk, we assume that C$^{18}$O, the most optically thin, best traces the CO column density. Because the millimeter continuum emission from the AS 209 disk is optically thin at all radii \citep{2012ApJ...760L..17P,2016AA...588A..53T}, dust opacity is expected to have negligible effects on the C$^{18}$O emission. Our CO surface density model is therefore motivated by the continuum-subtracted C$^{18}$O radial intensity profile. 

The gas surface density of protoplanetary disks is often modeled with the Lynden-Bell and Pringle \citeyearpar{1974MNRAS.168..603L} similarity solution for a viscous accretion disk. We adapt this for our model CO surface density profile (in cylindrical coordinates) by adding a Gaussian ring in the outer disk to simulate a large-scale return of CO into the gas phase: 
\begin{multline}
\Sigma_\text{CO}(r)  =\\ 
 \Sigma_c  \left  ( \left(\frac{r}{r_c}\right)^{-\gamma}\exp\left(-\left(\frac{r}{r_c}\right)^{2-\gamma}\right) + B\exp\left[-\frac{1}{2} \left( \frac{r-r_\text{ring}}{\sigma_\text{ring}} \right)^2\right]  \right)
\end{multline}

Following standard prescriptions for gas abundance distributions (e.g. \citealt{2013ApJ...775..136R}), we scale CO abundances vertically with the midplane pressure scale height:
\begin{equation}\label{eq:1}
\rho_\text{CO}(r,z) = \frac{\Sigma_\text{CO}(r)}{\sqrt{2\pi} H_\text{mid}(r)}\exp\left[-0.5\left(\frac{ z} {H_\text{mid}(r) }\right)^2\right],
\end{equation} 
with
\begin{equation}\label{eq:3}
H_\text{mid}(r) = \sqrt{\frac{k_B T_\text{mid}(r)r^3}{\mu_\text{gas}m_\text{H} G M_\ast}},
\end{equation}
where  $m_\text{H}$ is the mass of atomic hydrogen, and $\mu_\text{gas}=2.37$ is the mean gas particle mass.  

Similarly to Rosenfeld et al. \citeyearpar{2013ApJ...774...16R} and Dartois et al. \citeyearpar{2003AA...399..773D}, we model the vertical temperature gradient in cylindrical coordinates as  
\begin{equation}
T(r,z) =  \begin{cases} 
      T_{\textup{atm}}(r)+(T_{\textup{mid}}(r)-T_{\textup{atm}}(r))\cos^2{\left(\frac{\pi z}{2 z_q}\right)} & z\leq z_q \\
     T_{\textup{atm}}(r) & z>z_q 
   \end{cases}
\end{equation}
where
\begin{equation}
T_\textup{atm}(r) = T_{\textup{atm},10}\left(\frac{r}{\textup{10 AU}}\right)^{-q_\textup{atm}}
\end{equation}
\begin{equation}
T_\textup{mid}(r) = T_{\textup{mid},10}\left(\frac{r}{\textup{10 AU}}\right)^{-q_\textup{mid}}.
\end{equation}
$ T_{\textup{atm},10}$ and $T_{\textup{mid},10}$ are the atmosphere and midplane temperatures, respectively, at $r = 10$ AU. The temperature gradient scale height $z_q$ is  
\begin{equation}
\label{eq:6}
z_q = 4H_\text{mid}(r). 
\end{equation}
From a power-law fit to the midplane dust temperature that Andrews et al. \citeyearpar{2009ApJ...700.1502A} calculated for the AS 209 disk, $T_\text{10,mid} =47.3$ K and $q_\text{mid}$ = 0.48.  Since the atmosphere gas temperature profile of the AS 209 disk has not been constrained, we set $q_\textup{atm} = 0.5$ and $T_{\textup{atm},10} = 150$ K, similar to values Dartois et al. \citeyearpar{2003AA...399..773D} found for the disk around T Tauri star DM Tau. After some experimentation to determine parameter values that could reproduce the main features of the observations, we set $\Sigma_c = 5\times 10^{-5} \text{ g cm}^{-2}$, $\gamma = 1$, $r_c = 100$ AU, $B = 3$, $r_\text{ring} = 150$ AU, and $\sigma_\text{ring} = 20$ AU. 

\begin{figure*}[htp]
\epsscale{1.15}
\plotone{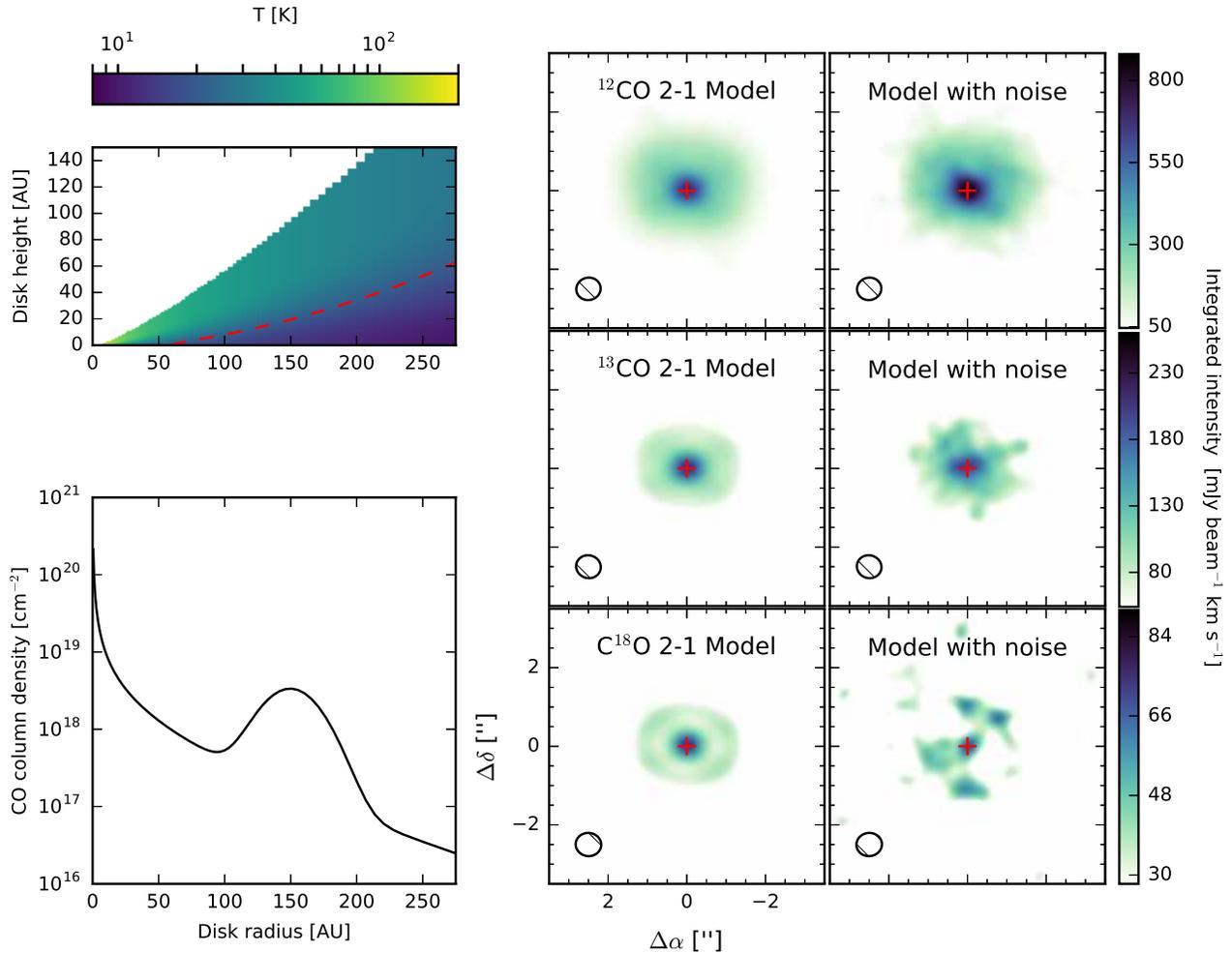}
\caption{Model structures and resulting integrated intensity maps for the model with an outer CO ring ($B= 3$ in eq. \ref{eq:1}). Top left: Temperature gradient profile up to four pressure scale heights (Eq. \ref{eq:3}) above the midplane. Red line denotes 20 K isotherm. Bottom left: CO column density (bottom). Middle: Integrated intensity maps for $^{12}$CO, $^{13}$CO, and C$^{18}$O $J=2-1$. Right column: Corresponding maps with gaussian noise added to model visibilities. Synthesized beams are drawn in the lower left. Red crosses mark phase centers. 
\label{fig3}}
\end{figure*}

\begin{figure*}[htp]
\epsscale{1.15}
\plotone{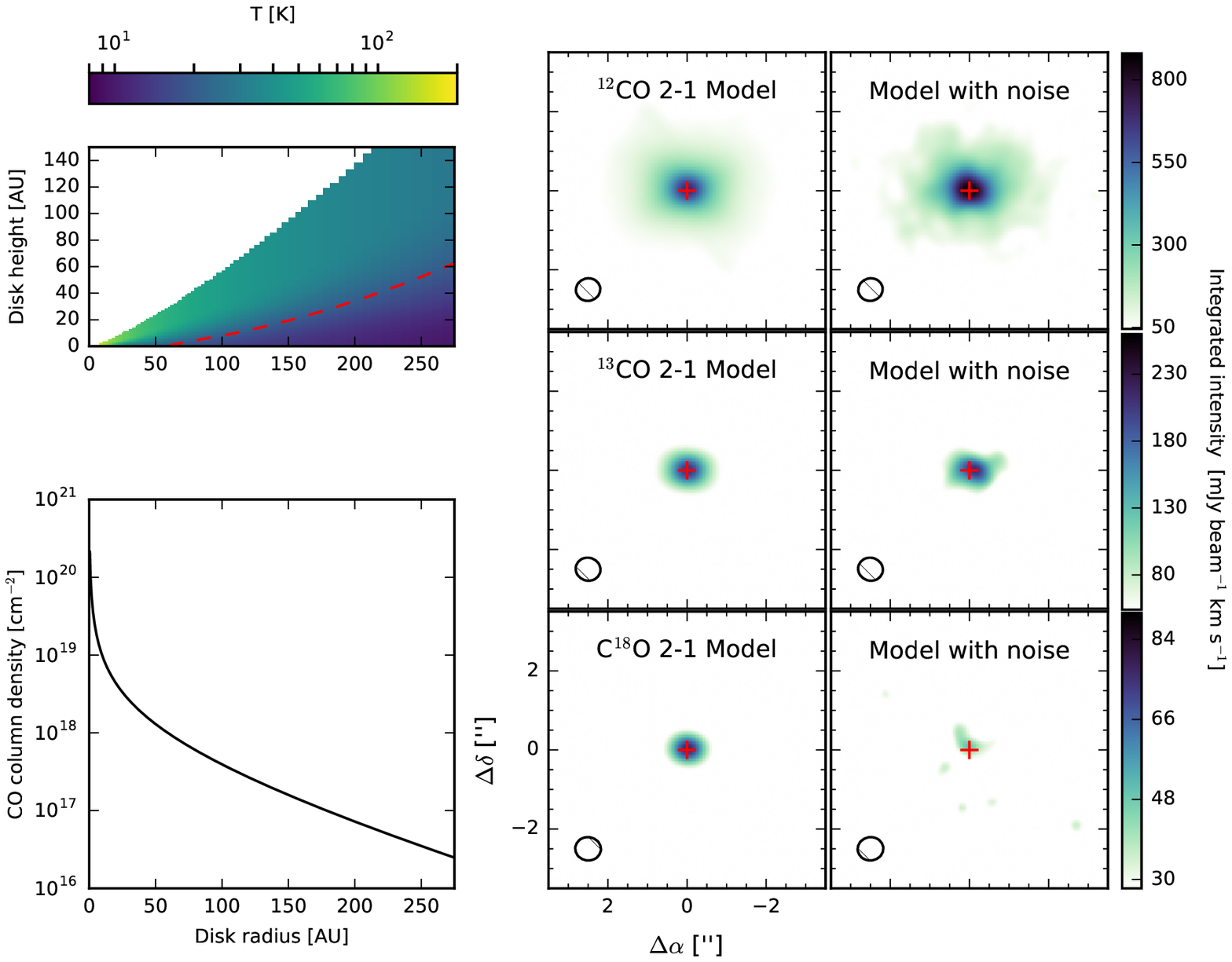}
\caption{Same as Figure \ref{fig3}, but for a model with monotonically decreasing CO column densities ($B= 0$ in eq. \ref{eq:1}). 
\label{fig4}}
\end{figure*}

 We use this model and the LAMDA database molecular data \citep{2005AA...432..369S, 2010ApJ...718.1062Y} to calculate $^{12}$CO, $^{13}$CO, and C$^{18}$O $J=2-1$ intensities with the radiative transfer code \textsc{RADMC-3D} \citep{2012ascl.soft02015D}. We fix the $^{12}$CO/$^{13}$CO and $^{12}$CO/C$^{18}$O ratios to the ISM values of 69 and 557, respectively \citep{1999RPPh...62..143W}. Local thermal equilibrium was assumed because CO and its isotopologues have low critical densities relative to disk gas densities \citep{2007ApJ...669.1262P}. Keplerian gas velocities were assumed, and the turbulent broadening parameter was set to $\xi =$ 0.01 km s$^{-1}$, on par with the TW Hya disk upper limit Hughes et al. \citeyearpar{2011ApJ...727...85H} derived. After computing sky brightness images at the orientation of the AS 209 disk, model visibilities were produced at the same spatial frequencies as the AS 209 data by using Python package \textsc{vis\_sample}.\footnote{Available at \url{https://github.com/AstroChem/vis_sample}} Model visibilities were imaged and CLEANed in CASA 4.4.0, then integrated across the same velocity ranges as the observations to produce integrated intensity maps. Observations were then simulated by adding Gaussian noise to the model visibilities so that the rms in the model image cubes was comparable to that of the data image cubes. Temperature and column density profiles and integrated intensity maps are shown for the model with the Gaussian CO ring in Figure \ref{fig3}. For comparison, we also model CO abundances without an outer ring by setting $B = 0$ in Eq. \ref{eq:1}, holding other parameters the same. Corresponding structure plots and integrated intensity maps are shown in Figure \ref{fig4}.  

For either model, the $^{12}$CO integrated emission is centrally peaked, in line with observations. However, only the model with an outer CO ring, shown in Figure \ref{fig3}, reproduces the central peak and ring in the C$^{18}$O emission, and the shoulder in $^{13}$CO emission. The morphological similarities between the data and model with an outer CO ring suggest that excess CO in the outer disk in conjunction with line opacity can account for the variation in emission patterns for the AS 209 disk's CO isotopologues. The model results highlight that a non-monotonic CO radial column density profile can be obscured in (partially) thick $^{12}$CO and $^{13}$CO observations, and thus direct evidence of CO desorption in the outer disk would often require observations of rarer, optically thin isotopologues such as C$^{18}$O.

\section{Discussion}
\subsection{Origin of the CO isotopologue emission morphology}
The outer emission ring in C$^{18}$O observed in the AS 209 disk suggests that its CO abundances are enhanced at large radii. In the radial intensity profile of C$^{18}$O in Figure \ref{fig1}, the ring peaks at $\sim$ 120 AU. While no snowline estimate has been reported in the literature for the AS 209 disk, the CO snowline has been estimated to lie at 30 AU for the TW Hya disk and 90 AU for the HD 163296 disk, corresponding to midplane temperatures of 17 and 25 K, respectively \citep{2013Sci...341..630Q, 2015ApJ...813..128Q}. TW Hya is a T Tauri star with $T_\text{eff}= 3400$\textendash4000 K and $ L_\ast$ = 0.19\textendash 0.23 $L_\odot$ \citep{1999ApJ...512L..63W,2011ApJ...732....8V}. AS 209 is somewhat warmer ($T_\text{eff} = 4250$ K and $L_\ast = 1.5$ $L_\odot$), suggesting that its CO snowline should lie at a slightly larger radius compared to TW Hya. As a Herbig Ae star, HD 163296 has a much warmer disk, so its CO snowline radius  can be treated as an upper bound for AS 209. Furthermore, the AS 209 disk midplane dust temperature profile derived by Andrews et al. \citeyearpar{2009ApJ...700.1502A} drops to 25 K at 35 AU and 17 K at 80 AU, suggesting that the CO snowline lies within this range. The decrement in emission between the central peak and outer ring of C$^{18}$O $J=2-1$ in the AS 209 disk occurs at $\sim0 \farcs 7$  (80 AU), consistent with the expected location of initial CO freezeout. Meanwhile, the C$^{18}$O ring in the AS 209 disk occurs well outside the expected CO snowline.  

Based on T Tauri disk models indicating that radial drift could increase midplane dust temperatures at large radii, Cleeves \citeyearpar{2016ApJ...816L..21C} suggested that subsequent thermal desorption of CO in the outer disk could cause CO enhancement followed by a second CO snowline once temperatures began falling again. P\'erez et al. \citeyearpar{2012ApJ...760L..17P} found that larger dust grains emitted from a more compact region than small grains in the AS 209 disk, providing evidence of radial drift of solids. Cleeves' models and P\'erez et al.'s observations suggest that the C$^{18}$O emission morphology in the AS 209 disk can plausibly be explained by thermal CO desorption in the outer disk, with CO snowlines both interior and exterior to the desorption ring.  Non-thermal desorption may also substantially enhance outer disk CO. Further observations and modeling are necessary to constrain the relative contributions of thermal versus non-thermal processes.     

Simulated observations from our toy model of CO desorption in the outer disk indicate that isotopologue opacity differences in the AS 209 disk can explain why there is a C$^{18}$O ring, but only a shoulder in $^{13}$CO and a monotonically decreasing intensity profile in $^{12}$CO. Selective photodissociation is a less likely alternative. Models from \citet{2014AA...572A..96M} indicate that selective photodissociation leads to lower C$^{18}$O intensities in the outer disk compared to models without, whereas we observe an increase in C$^{18}$O intensity in the outer disk of AS 209. 

Another possibility to examine is that dust opacity creates the appearance of a ring in optically thin C$^{18}$O emission. The peak brightness temperature of the 1.4 mm AS 209 continuum is only 7 K, which is consistent with the findings from \citealt{2012ApJ...760L..17P} and \citealt{2016AA...588A..53T} of an optically thin disk. However, \citealt{2015ApJ...810..112O} suggested that such low brightness temperatures could also be consistent with dust concentrated into optically thick rings much narrower than the beam. While existing observations do not point to this scenario, it could be tested with ALMA at higher spatial resolutions.   

\subsection{Implications for disk structure modeling}
Disk structure inferences made from CO observations depend on assumptions about chemistry (e.g. \citealt{2013ApJ...776L..38F, 2014AA...572A..96M,2014ApJ...788...59W}). Common parametric disk gas structure models stipulate that CO emission decreases monotonically with radius \citep{2013ApJ...774...16R,2013ApJ...775..136R}, in line with most $^{12}$CO observations and chemical models \citep{2001AA...371.1107A, 2007ApJ...660..441W,2012ApJ...747..114W}. However, the optically thick lower-$J$ transitions of $^{12}$CO in disks could easily obscure CO substructures, as indicated by our toy model. Although CO isotopologue emission rings have been observed in disks, they have been primarily associated with systems with prominent dust cavities, suggesting that gas and dust cleared for similar reasons \citep{2014AA...562A..26B, 2016AA...585A..58V,2016ApJ...820...19T}. The non-monotonicity of the C$^{18}$O radial intensity profile for the AS 209 disk, as well as the double DCO$^+$ rings of the IM Lup disk in \citet{2015ApJ...810..112O}, are unexpected because neither disk shows millimeter dust gaps. Since the outer  C$^{18}$O emission ring observed in the AS 209 disk seems inconsistent with the standard parameterizations used to model CO emission in disks, the observations presented in this paper indicate that desorption effects should also be accounted for when tracing disk structure with CO. 

\subsection{Developing further constraints on desorption}
Given the evidence of CO desorption in the AS 209 disk in this work, as well as observations of IM Lup \citep{2015ApJ...810..112O}, we expect that signatures of CO desorption and second snowlines will be apparent in the outer regions of other disks observed at high resolution with ALMA. Recent observations of $^{13}$CO and C$^{18}$O $J = 3-2$ in the TW Hya disk indicate an outer ring, which \citet{schwarz} suggested could be due to desorption. Because optically thick $^{12}$CO and even $^{13}$CO transitions could obscure significant features in CO abundance profiles, especially for more massive, gas-rich disks, sensitive high-resolution observations of C$^{18}$O and C$^{17}$O can place further constraints on desorption mechanisms and better inform CO emission parameterizations. {\"O}berg et al.  \citeyearpar{2015ApJ...810..112O} suggested that observing multiple transitions would provide better estimates of gas temperature in the outer disk, thereby distinguishing between thermal and non-thermal desorption. Better temperature measurements will also be necessary to derive CO column densities more reliably in the outer disk. Furthermore, N$_2$H$^+$ is often used to locate the CO snowline because its formation is inhibited and its destruction accelerated by gas-phase CO \citep{2013Sci...341..630Q}. If abundant CO returns to the gas phase near the midplane, then N$_2$H$^+$ may trace this desorption front.   

\subsection{Summary}
We present ALMA observations of $^{12}$CO, $^{13}$CO, and C$^{18}$O $J=2-1$ emission in the AS 209 disk. The C$^{18}$O emission ring is interpreted as the onset of large-scale CO desorption in the outer disk. Based on simulated observations from a toy model, we propose that CO desorption in the outer disk, along with line opacity differences, can explain variations in emission patterns among the isotopologues. The CO isotopologue emission in AS 209 is not described well by common parametric models used to infer disk structure from CO observations, indicating that CO is not a straightforward tracer of total disk gas properties. To constrain the effects of desorption on CO abundances, especially on the creation of outer disk CO desorption fronts or secondary snowlines, high-resolution observations of rare, optically thin CO isotopologues are crucial.   

\acknowledgments

We thank the referee for comments improving this paper, Ian Czekala, Ilse Cleeves, and Meredith MacGregor for helpful discussions, and Ryan Loomis for access to \textsc{vis\_sample}. This paper makes use of ALMA data ADS/JAO.ALMA\#2013.1.00226.S. ALMA is a partnership of ESO (representing its member states), NSF (USA) and NINS (Japan), together with NRC (Canada) and NSC and ASIAA (Taiwan), in cooperation with the Republic of Chile. The Joint ALMA Observatory is operated by ESO, AUI/NRAO and NAOJ. The National Radio Astronomy Observatory is a facility of the National Science Foundation operated under cooperative agreement by Associated Universities, Inc.  KI\"O acknowledges funding from the Alfred P. Sloan Foundation and the Packard Foundation. This material is based upon work supported by the National Science Foundation Graduate Research Fellowship under Grant No. DGE-1144152. 

\software{CASA 4.4.0, RADMC-3D, \textsc{pwkit} (\url{https://github.com/pkgw/pwkit}), \textsc{vis\_sample} (\url{https://github.com/AstroChem/vis_sample}), \textsc{cubehelix} (\url{https://github.com/jradavenport/cubehelix})}

\end{document}